\documentclass[manuscript]{aastex}

\usepackage{epsfig}
\usepackage[]{natbib}

\shorttitle{IPN Supplement}
\shortauthors{Hurley et al.}

\begin{document}

\bibliographystyle{plainnat}

\title{The Interplanetary Network Supplement to the BeppoSAX Gamma-Ray Burst Catalogs}

\author{K. Hurley}
\affil{Space Sciences Laboratory, University of California,
7 Gauss Way, Berkeley, CA 94720-7450, U.S.A.}
\email{khurley@ssl.berkeley.edu}

\author{C. Guidorzi, F. Frontera\altaffilmark{1}, E. Montanari\altaffilmark{2}, F. Rossi}
\affil{University of Ferrara, Physics Department, Via Saragat, 1, 44100 Ferrara, Italy}

\altaffiltext{1}{INAF/Istituto di Astrofisica Spaziale e Fisica Cosmica di Bologna, via Gobetti 101, I-40129 Bologna, Italy}
\altaffiltext{2}{Istituto IS Calvi, Finale Emilia (MO), Italy}

\author{M. Feroci}
\affil{INAF - Istituto di Astrofisica Spaziale e Fisica Cosmica, via Fosso del Cavaliere, Rome, I-00133, Italy}

\author{E. Mazets, S. Golenetskii, D. D. Frederiks, V. D. Pal'shin, R. L. Aptekar}
\affil{Ioffe Physico-Technical Institute of the Russian
Academy of Sciences, St. Petersburg, 194021, Russian Federation}

\author{T. Cline\altaffilmark{3}, J. Trombka, T. McClanahan, R. Starr}
\affil{NASA Goddard Space Flight Center, Greenbelt, MD 20771, U.S.A.}
\altaffiltext{3}{Emeritus}

\author{J.-L. Atteia, C. Barraud, A. P\'{e}langeon}
\affil{Laboratoire d'Astrophysique, Observatoire Midi-Pyr\'{e}r\'{e}es,
14 avenue E. Belin, 31400 Toulouse, France}

\author{M. Bo\"{e}r}
\affil{Observatoire de Haute-Provence, 04870 Saint Michel l'Observatoire, France}

\author{R. Vanderspek, G. Ricker}
\affil{Kavli Institute for Astrophysics and Space Research, Massachusetts Institute of Technology,
70 Vassar Street, Cambridge, MA 02139, U.S.A.}

\author{I. G. Mitrofanov, D. V. Golovin, A. S. Kozyrev, M. L. Litvak, A. B. Sanin}
\affil{Space Research Institute, 84/32, Profsoyuznaya, Moscow 117997, Russian Federation}

\author{W. Boynton, C. Fellows, K. Harshman}
\affil{University of Arizona, Department of Planetary Sciences, Tucson, Arizona 85721, U.S.A.}

\author{J. Goldsten, R. Gold}
\affil{Applied Physics Laboratory, Johns Hopkins University, Laurel, MD 20723, U.S.A.}

\author{D.M. Smith}
\affil{Physics Department and Santa Cruz Institute for Particle Physics,
University of California, Santa Cruz, Santa Cruz, CA 95064, U.S.A.}

\author{C. Wigger, W. Hajdas}
\affil{Paul Scherrer Institute, 5232 Villigen PSI, Switzerland}

\bf Date: 8 April 2010 \rm

\begin{abstract}
Between 1996 July and 2002 April, one or more spacecraft of the interplanetary
network detected 787 cosmic gamma-ray bursts that were also detected
by the Gamma-Ray Burst Monitor and/or Wide-Field X-Ray Camera experiments aboard the \it BeppoSAX \rm spacecraft.  During this period, the 
network consisted of up to six spacecraft,
and using triangulation, the localizations of 475 bursts were obtained.
We present the localization data for these events.
\end{abstract}


\section{Introduction}

Between 1996 July and 2002 April, the Wide Field X-Ray Camera (WFC) and Gamma-Ray Burst Monitor (GRBM) aboard the \it BeppoSAX \rm
mission detected 62 and 1092 cosmic gamma-ray bursts, respectively, and localized many of them  to accuracies
which ranged from arcminutes to tens of degrees \citep{vetere07,frontera09}; instrument descriptions may be found
in \citet{feroci97}, \citet{frontera97}, and \citet{jager97}.  These detections were used
to initiate searches through the data of the spacecraft comprising the interplanetary
network (IPN). In 475 cases localizations could be obtained by
triangulation, and successful multiwavelength counterpart searches were initiated for some of them.  
The IPN contained between 4 and 6 spacecraft during this period.  They were,
in addition to \it BeppoSAX \rm:
\it Ulysses, \rm in heliocentric orbit at distances between 670 and 3180 
light-seconds from Earth \citep{hurley92}; \it Konus-Wind \rm, in various orbits up to around
4 light-seconds from Earth \citep{aptekar95}; \it HETE-II - FREGATE \rm , in low Earth orbit \citep{ricker03,atteia03}; the \it Near-Earth 
Asteroid Rendezvous \rm mission (NEAR), 
at distances up to 1300 light-seconds 
from Earth \citep{trombka99}; \it Mars Odyssey\rm, launched in 2001 April and in orbit around Mars starting in
2001 October, up to 1250 light-seconds from Earth \citep{hurley06a}; the
\it Compton Gamma-Ray Observatory \rm (the Burst and Transient Source Experiment, BATSE - \citet{fishman92}; and
\rm the \it Ramaty High Energy Solar Spectroscopic Imager \rm (RHESSI) both in low Earth orbit \citep{smith02}.
 Their timelines are presented in
figure 1.  In this paper, we present the localization data obtained by the IPN for these bursts.

At least three other spacecraft recorded GRB detections during this period, although they were not used
for triangulation and therefore were not, strictly speaking, part of the IPN.  The \it Rossi X-Ray Timing Explorer \rm (RXTE)  All Sky Monitor
detected and localized some \it BeppoSAX \rm bursts \citep{smith99}.  It operated in the low energy X-ray range,
where the light curves of gamma-ray bursts differ significantly from the high energy range where the other IPN instruments
operate.  The \it Defense Meteorological Satellite Program \rm (DMSP) \citep{terrell96, terrell98, terrell04} 
and the \it Stretched Rohini Satellite Series \rm (SROSS) \citep{marar94} spacecraft
detected, but did not localize bursts.

\section{Observations}

For each gamma-ray burst detected by
\it BeppoSAX \rm, a search was initiated in
the data of the IPN spacecraft.  For the spacecraft within a few light-seconds
of Earth, the search window was centered on the \it BeppoSAX \rm trigger time, and its
duration was somewhat greater than the event duration.
For the spacecraft at interplanetary distances, the search window was 
twice the light-travel time to the spacecraft if the event arrival direction
was unknown, which was the case for most events.  If the arrival direction
was known, even coarsely, the search window was defined by calculating the expected arrival time
at the spacecraft, and searching in a window around it.  
Of the approximately 3300 events detected by one or more IPN spacecraft while \it BeppoSAX \rm was operational, 787
were also detected by \it BeppoSAX \rm; these are listed in table 1, with the following abbreviations:
DMS: \it Defense Meteorological Satellite Program\rm, HET: \it HETE-II\rm, Kon: \it Konus-Wind\rm,
MO: \it Mars Odyssey\rm, NEA: \it Near Earth Asteroid Rendezvous \rm mission, RHE: \it Ramaty High Energy Solar Spectroscopic Imager\rm,
SRS: \it Stretched Rohini Satellite Series\rm, Uly: \it Ulysses\rm, XTE: \it Rossi X-Ray Timing Explorer\rm.  
Table 2 shows the number of events observed by each spacecraft in the IPN,
and table 3 gives the number of bursts that were detected by a total
of N spacecraft, where N is between 2 and 6 (detections by RXTE, 
DMSP, and SROSS have been counted).  

\section{Localizations}

When a GRB arrives at two spacecraft with a delay $\rm \delta$T, it may be
localized to an annulus whose half-angle $\rm \theta$ with respect to the
vector joining the two spacecraft is given by 
\begin{equation}
cos \theta=\frac{c \delta T}{D}
\end{equation}
where c is the speed of light and D is the distance between the two
spacecraft.  (This assumes that the burst is a plane wave, i.e. that its
distance is much greater than D.)  The annulus width d$\rm \theta$, is
\begin{equation}
d \rm \theta =c \rm \sigma(\delta T)/Dsin\rm \theta
\end{equation}
where
$\rm \sigma(\delta$T) is the uncertainty in the time delay.  
$\rm \sigma(\delta$T) is generally of the order of 100 ms or more, when
both statistical and systematic uncertainties are considered; thus triangulation
between two near-Earth spacecraft, for which D/c is at most $\sim$40 ms, does
not constrain the burst arrival direction significantly.  When D/c is of the order of
several light-seconds (e.g., the distance between \it Konus-Wind\rm \, and
a near-Earth spacecraft), annuli with widths of several degrees or less can be
obtained; when D/c is several hundred light-seconds or more (i.e. an interplanetary
spacecraft and a near-Earth spacecraft), annulus widths of the order of
arcminutes or less are possible.  When two interplanetary spacecraft and a near-Earth
spacecraft observe a GRB, a small error box can be obtained.  Table 4 gives
the number of events observed by 0, 1, and 2 interplanetary spacecraft.

475 bursts could be localized by the method above; table 5 gives the localization information
for them.  Triangulation annuli are
given in the 4 IPN columns: these are the right ascension and declination
of the annulus center $\alpha, \delta$, the annulus radius R, and the uncertainty in the
radius $\delta$R.  One or two annuli are specified.  In addition to triangulation
annuli, several other types of localization information
are included in this catalog.  The 3 BATSE columns give the right ascension,
declination, and 1 $\sigma$ (statistical only) error radius of the BATSE localizations.  
These are taken from the current catalog on the BATSE website (http://www.batse.msfc.nasa.gov/batse/grb/catalog/current/),
as well as from the BATSE untriggered burst catalogs \citep{stern01, kommers00}.
3 SAX columns give the right ascension, declination,
and 90 \% confidence radius of the \it BeppoSAX \rm localization, either from the GRBM \citep{frontera09} or
the WFC \citep{vetere07}, or from the IAU and GCN Circulars. 
The 3 HETE columns give the right ascension, declination, and radius of the Wide Field
X-Ray Monitor error circle 
(Vanderspek et al. 2010).  Combining these
error circles with the IPN annuli often results in smaller error regions.  
IPN localizations for almost all bursts with a BATSE or HETE error circle 
have appeared in a previous catalog and are repeated here only for completeness.

The two Ecliptic columns give the ecliptic latitudes of the bursts, measured
northward (positive) from the ecliptic plane towards the north ecliptic pole.  These are derived by comparing the count
rates of the two \it Konus-Wind\rm \, detectors (Aptekar et al. 1995).  Planet-blocking
is specified by the right ascension and declination of the planet's center and
its radius, in the 3 Planet columns.  When a spacecraft in low Earth or Mars
orbit observes a burst, the planet blocks up to $\approx$ 3.7 sr of the sky.
This is often useful for deciding which of two annulus intersections
is the correct one, or for eliminating portions of a single annulus.
Finally, the Other column gives the right ascension, declination, and
radius of any other localization region, which may be obtained in one of
several ways.  In some cases, the burst was observed by four spacecraft which
were separated by large enough distances to give 3 triangulation annuli, whose
intersections are consistent with a single error box.  In other cases, the
anisotropic response of one of the IPN experiments allows the ambiguity to
be resolved.  In still other cases, a region 
may be derived from planet-blocking by a second spacecraft in
addition to the data in the Planet column. In this case the
error circle given is the complement of the planet-blocking circle, that
is, a circle whose RA is the RA of the planet plus 180 degrees, whose
declination is the negative of the planet's declination, and whose radius
is 180 degrees minus the planet's angular radius.    
The units of all entries in table 5 are degrees, and all coordinates are J2000.  
For some events, no triangulation was possible, but coarse constraints on
the burst arrival direction can
be derived from planet-blocking, ecliptic latitudes, or both.  This information
is not given here, but information
on these events, as well as the ones in this catalog, may be found at the
IPN website: \url{ssl.berkeley.edu/ipn3/index.html }. 
Figures 2 and 3 show examples of coarse and fine IPN localizations.

As for BATSE, the \it BeppoSAX \rm GRBM localizations are derived by comparing the count rates
of various detectors aboard these spacecraft.  These localizations are affected by Earth albedo and absorption
by spacecraft materials, among other things, and their shapes are in general complex.  The error
circles are approximations to these shapes.  They are centered at the point which is
the most likely arrival direction for the burst, and their radii are defined so that their
areas are equal to the 1 $\sigma$ (BATSE) or 90 \% confidence (\it BeppoSAX \rm GRBM) statistical-only true
error regions.  Therefore in some cases, indicated by a footnote, the IPN annuli do not cross the error circles.
This occurs for 25 of the 133 \it BeppoSAX \rm GRBM localizations in this catalog.  We have examined the true \it BeppoSAX \rm error regions in all of these
cases and have verified that they are indeed consistent with the IPN annuli.  In some of these cases, an error circle has been
defined in the ``Other'' column which limits the IPN annulus or annuli to a region which, from a consideration
of all the available data, is known to define the arrival direction.  Thus for those bursts where the GRBM error
circle does not intersect the IPN annulus, the ``Other'' circle should be used in place of the GRBM circle.  

Table 6 gives the approximate localization area in square 
degrees for each of the bursts in table 5.  This is the area of the region which is common to all the localizations
given in table 5.  For bursts where the \it BeppoSAX \rm or BATSE error circle does not intersect the IPN annulus,
the area given is that of the annulus alone.

\section{Comments on specific events}

GRB960916 at 03:56:20 is GRB960916B in the \it BeppoSAX \rm catalog \citep{frontera09}.  GRB960916A
occurred 312 s earlier, at 03:51:08, and it was detected by \it Konus-Wind \rm, but not by \it Ulysses \rm.
This non-detection is consistent with the fact that the earlier event was weaker.
The \it Konus \rm ecliptic latitudes for these two events are consistent with a single origin, i.e.
a very long burst.

GRB970315 at 22:09:19 (GRB970315B in \citet{frontera09} may be from the same source as
BATSE 6125 at 22:13:42 (http://www.batse.msfc.nasa.gov/batse/grb/catalog/current/).  The IPN annulus
passes through the BATSE error circle, and the duration of the BATSE event is given as 1307 s.
BeppoSAX entered the SAA at 22:10:09, so it could not observe the BATSE event, and the BATSE position
of the event was Earth-occulted to BATSE at the time of the BeppoSAX event.  If these are indeed from
a single source, the total duration would have been around 1570 s.  \it Ulysses \rm did not observe
any emission which would be consistent with the BATSE burst, but this is consistent with its lower
intensity.

GRB970415 was observed as a very weak event by \it Ulysses \rm, and reliable triangulation of it
is not possible. 

GRB970518 has a duration of approximately 370 s.  The GRBM observed only the later part of the event, at
07:12:12.  However, the burst started at 07:06:23, and this is the time given in tables 1, 5, and 6.

GRB971228B at 14:53:52 was observed as a very weak event by \it Ulysses \rm, and reliable triangulation of it
is not possible.

GRB990516A at 20:55:15 was observed as a very weak event by \it Ulysses \rm, and reliable triangulation of it
is not possible.

GRB990905 at 22:38:55 was observed as a very weak event by \it Ulysses \rm, and reliable triangulation of it
is not possible.

GRB991026 has an IPN localization which is inconsistent with the final \it BeppoSAX \rm WFC localization in \citet{vetere07}.
The minimum distance between the IPN annulus and the WFC position is about 4.8 degrees
(no uncertainty is given for the WFC localization).  
The WFC position given is from in't Zand (private communication, 2004).

GRB991030 has an IPN localization which is inconsistent with the \it BeppoSAX \rm WFC localization in \citet{vetere07}.
The minimum distance between the IPN annulus and the WFC position is about 5.9 degrees
(no uncertainty is given for WFC localization).  The WFC position given is from in't Zand (private communication, 2004).

GRB000629 does not appear in the \it BeppoSAX \rm catalog, because it was initially thought to be solar.
Analysis of the Konus-Wind data, however, points to a likely cosmic origin.

GRB011221 triggered the GRBM just prior to entry into the South Atlantic Anomaly.  All GRBM data were lost,
and this burst does not appear in the \it BeppoSAX \rm catalog.

\section{Conclusions}

This is the eleventh in a continuing series of IPN catalogs \citep{hurley99a, hurley99b, hurley00a, hurley00b,
hurley00c, hurley05, hurley06b, laros97, laros98,
hurley09}; the localization
data for all of them can be found in electronic form at the IPN website.
The IPN is, in effect, a full-time, all-sky monitor, when the duty cycles
and viewing constraints of all its instruments are considered.  Its fluence and flux thresholds
for 50\% detection efficiency are about $\rm 6 \times 10^{-7} erg \, cm^{-2}\, and \,
1 \, photon \, cm^{-2}\, s^{-1}$, \,respectively.  Over the \it BeppoSAX \rm mission, 787 bursts were detected by
the GRBM and/or the WFC and at least one other IPN instrument and 475 of them could be localized
to some extent by triangulation.  The more precise and/or rapid localizations were announced in
over 50 IAU and GCN Circulars (in 1997, and in 1998 -- 2002, respectively), 
resulting in multiwavelength counterpart searches.
Regardless of precision and speed of the localizations, however, burst data such as this are useful for numerous studies,
such as searching for indications of activity from previously unknown soft gamma repeaters,
associating supernovae with bursts, or searching for neutrino and gravitational radiation
associated with bursts. 

\section{Acknowledgments}

Support for the interplanetary network came from the following sources:
JPL Contracts 958056 and 1268385 (Ulysses); MIT Contract SC-R-293291 and
NASA NAG5-11451 (HETE); NASA NNX07AH52G (Konus); NASA NAG5-13080 (RHESSI); 
NASA NAG5-11451 and JPL Contract 1282043 (Odyssey);  NASA NAG5-7766, NAG5-9126, NAG5-10710, 
and the U.S. SAX Guest Investigator program (\it BeppoSAX \rm); 
and NASA NAG5-9503 (NEAR).  C.G., F.F., and E.M. acknowledge financial support from the ASI-INAF
contract I/088/06/0.  In Russia,
this work was supported by the Federal Space Agency of Russia and RFBR grant 09-02-00166a.

\begin{figure}
\epsscale{.5}
\includegraphics[angle=0,scale=.5]{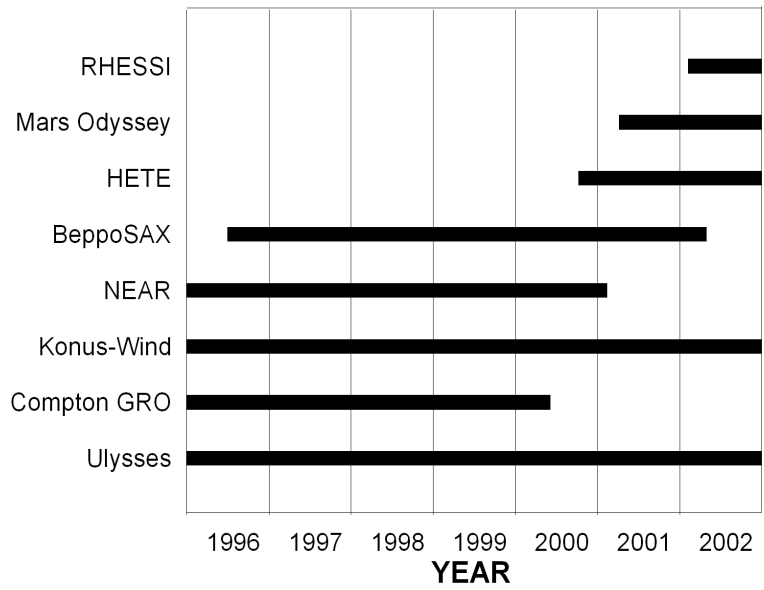}
\caption{The timelines of the missions comprising the interplanetary network
between 1996 and 2002.  During the period when \it BeppoSAX \rm was operational, there were a minimum of
3 and a maximum of 5 other missions in the network.  There were two interplanetary spacecraft
in operation for most of the \it BeppoSAX \rm mission, \it Ulysses \rm and either NEAR or \it Odyssey \rm.}\label{fig1}
\end{figure}

\begin{figure}{}
\epsscale{1}
\includegraphics[scale=1.50]{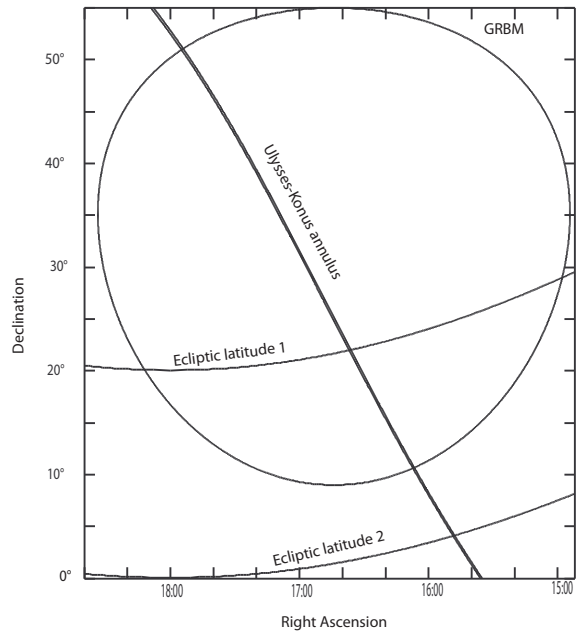}
\caption{Localizations of GRB970203.  The arrival direction is defined by the intersection of the 33 degree radius
GRBM error circle, the 0.16 degree wide IPN annulus, and the 20 degree wide Konus ecliptic latitude band.}\label{fig2}
\end{figure}

\begin{figure}{}
\epsscale{1}
\includegraphics[scale=.50]{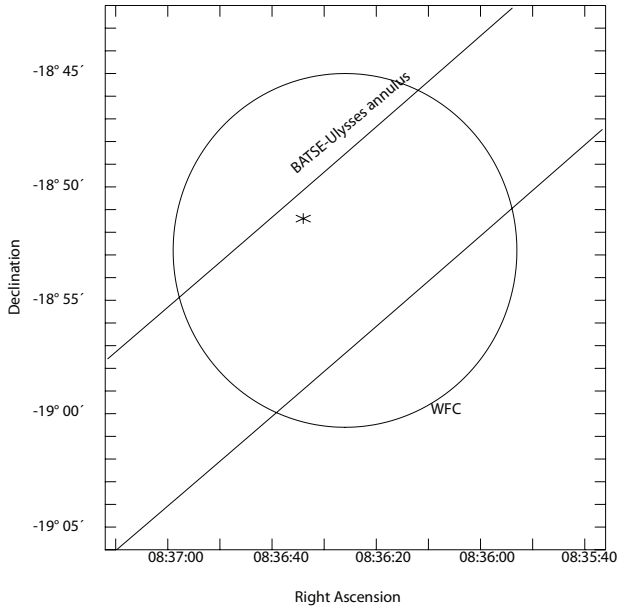}
\caption{Localizations of GRB980326.  The arrival direction is defined by the intersection of the 0.133 degree
radius WFC error circle and the .092 degree wide BATSE-Ulysses annulus.  The initial WFC and IPN localizations
were announced in Celidonio et al. (1998) and Hurley et al. (1998).  The optical counterpart, indicated by
an asterisk, was found by Groot et al. (1998).}\label{fig3}
\end{figure}

\clearpage

\begin{deluxetable}{cccc}
\tabletypesize{\scriptsize}
\tablecaption{IPN/\it BeppoSAX \rm gamma-ray bursts}
\tablewidth{0pt}
\tablehead{
\colhead{Date} & \colhead{Universal Time}\tablenotemark{a}  & 
\colhead{Observed by}\tablenotemark{b} 
}
\startdata

1996 Jul 03 & 07:39:48 & Kon \\
1996 Jul 03 & 13:42:53 & Uly,BATSE,Kon \\
1996 Jul 03 & 18:10:40 & BATSE,Kon,SRS \\
1996 Jul 07 & 10:16:40 & Uly,BATSE,Kon \\
1996 Jul 07 & 16:26:04 & BATSE \\
1996 Jul 20 & 11:36:53 & BATSE,Kon \\
1996 Jul 23 & 04:46:01 & BATSE,Kon \\
1996 Jul 25 & 17:39:07 & BATSE \\
\enddata

\tablenotetext{a}{Universal time is the Earth-crossing time of the start of the event}
\tablenotetext{b}{
DMS: \it Defense Meteorological Satellite Program\rm; 
HET: \it HETE-II\rm; 
Kon: \it Konus-Wind\rm; 
MO: \it Mars Odyssey\rm; 
NEA: \it Near Earth Asteroid Rendezvous \rm mission; 
RHE: \it Ramaty High Energy Solar Spectroscopic Imager\rm; 
SRS: \it Stretched Rohini Satellite Series\rm;
Uly: \it Ulysses\rm; 
XTE: \it Rossi X-Ray Timing Explorer \rm }
\end{deluxetable}

\clearpage

\begin{deluxetable}{cccccccccc}
\tablewidth{0pt}
\tabletypesize{\footnotesize}
\tablecaption{Number of \it BeppoSAX \rm bursts observed by each spacecraft}
\tablehead{
\colhead{BATSE} & \colhead{DMSP} & \colhead{HETE} & \colhead{Konus} & \colhead{NEAR} & \colhead{Odyssey} &  \colhead{RHESSI} &  \colhead{RXTE} & \colhead{SROSS} & \colhead{Ulysses} 
}
\startdata
         426    &           14    &           14   &          528    &          152   &            13     &           10      &            9    &           5     &           376     \\
\tableline
\enddata
\end{deluxetable}

\clearpage
\begin{deluxetable}{ccccccccc}
\tablewidth{0pt}
\tablecaption{Number of \it BeppoSAX \rm bursts in this catalog observed by a total of N experiments, regardless of their
distance from Earth}
\tablehead{
\colhead{N=1}           & \colhead{2}      &  \colhead{3}          & \colhead{4}  &  \colhead{5} & \colhead{6}    &
 }
\startdata
          0             &         324      &          220          &         196  &          40  &          7  \\
\enddata
\end{deluxetable}

\clearpage

\begin{deluxetable}{ccc}
\tablecaption{Number of \it BeppoSAX \rm bursts observed by N interplanetary spacecraft, i.e., NEAR, \it Mars Odyssey \rm, and
\it Ulysses \rm}
\tablewidth{0pt}
\tablehead{
\colhead{N=0}           & \colhead{1}      &   \colhead{2} }
\startdata
362                     & 309              & 116       \\
\enddata
\end{deluxetable}

\clearpage
\pagestyle{empty}

\begin{deluxetable}{ccccccccccccccccccccccc}
\setlength{\tabcolsep}{.05in}
\tabletypesize{\scriptsize}
\tablewidth{0pt}
\rotate
\tablecaption{IPN localization data}
\tablehead{
\multicolumn{2}{}{} & \multicolumn{3}{|c|}{BATSE} & \multicolumn{3}{|c|}{SAX} & \multicolumn{3}{|c|}{HETE} & \multicolumn{4}{|c|}{IPN} &
\multicolumn{2}{|c|}{Ecliptic} & \multicolumn{3}{|c|}{Planet} & \multicolumn{3}{|c|}{Other} \\
\colhead{Date} & \colhead{UT} & \colhead{$\alpha$} & \colhead{$\delta$} & \colhead{R} & \colhead{$\alpha$} & \colhead{$\delta$} & \colhead{R} & 
\colhead{$\alpha$} & \colhead{$\delta$} & \colhead{R} & \colhead{$\alpha_1$} & \colhead{$\delta_1$} & \colhead{$R_1$} & \colhead{$\delta R_1$} & 
\colhead{$\beta_1$} & \colhead{$\beta_2$} &
\colhead{$\alpha$} & \colhead{$\delta$} & \colhead{R} & \colhead{$\alpha$} & \colhead{$\delta$} & \colhead{R} \\
\colhead{}         & \colhead{}         &  \colhead{} & \colhead{}         &  \colhead{}        & \colhead{}      & \colhead{}      & \colhead{}   &
\colhead{}         & \colhead{}         &  \colhead{} & 
\colhead{$\alpha_2$} & \colhead{$\delta_2$} & \colhead{$R_2$} & \colhead{$\delta R_2$} &
\colhead{}           & \colhead{}           & \colhead{}      &  \colhead{}            &  \colhead{}              & \colhead{}      & \colhead{}   &  \colhead{}                 
}
\startdata

03 Jul 1996 & 13:42:53 &   4.6      &  -7.8      &   1.0      & ...        & ...        & ...        & ...        & ...        & ...        & 334.713       & -46.652       &  46.702       &    .136       & ...        & ...        & ...        & ...        & ...        & ...        & ...        & ...        \\
07 Jul 1996 & 10:16:40 & 321.0      &  82.5      &   1.1      & ...        & ...        & ...        & ...        & ...        & ...        & 155.435       &  45.862       &  50.112       &    .027       & ...        & ...        & ...        & ...        & ...        & ...        & ...        & ...        \\
31 Jul 1996 & 05:46:01 &  43.9      &  12.5      &    .9      & ...        & ...        & ...        & ...        & ...        & ...        & 340.144       & -41.403       &  83.839       &    .012       & ...        & ...        & ...        & ...        & ...        & ...        & ...        & ...        \\
01 Aug 1996 & 11:29:09 & ...        & ...        & ...        & ...        & ...        & ...        & ...        & ...        & ...        & 340.436       & -41.223       &  78.075       &    .099       &  -9.8      &  10.2      & ...        & ...        & ...        & ...        & ...        & ...        \\
05 Aug 1996 & 21:55:57 & ...        & ...        & ...        & ...        & ...        & ...        & ...        & ...        & ...        & 161.328       &  40.481       &  74.569       &    .012       &  33.8      &  53.8      & 223.8      &   3.8      &  66.1      & ...        & ...        & ...        \\
            &          &            &            &            &            &            &            &            &            &            & 161.301       &  40.450       &  74.588       &    .023       &            &            &            &            &            &            &            &            \\
\enddata
\tablenotetext{a}{IPN annulus does not include \it BeppoSAX \rm localization}
\end{deluxetable}


\begin{deluxetable}{cccc}
\tabletypesize{\scriptsize}
\tablecaption{Error box areas}
\tablewidth{0pt}
\tablehead{
\colhead{Date} & \colhead{Universal Time} & \colhead{Area, square degrees}  
}
\startdata

03 Jul 1996 & 13:42:53 & 1.09E+00 \\
07 Jul 1996 & 10:16:40 & 1.63E-01 \\
31 Jul 1996 & 05:46:01 & 9.28E+00 \\
01 Aug 1996 & 11:29:09 & 1.01E+01 \\
05 Aug 1996 & 21:55:57 & 5.83E-01 \\
10 Aug 1996 & 06:56:16 & 3.75E-02 \\
25 Aug 1996 & 00:23:03 & 1.49E+00 \\
12 Sep 1996 & 13:57:27 & 1.54E-01 \\
\enddata
\end{deluxetable}

\end{document}